\documentclass[10pt,letterpaper]{article}
\usepackage[top=0.85in,left=2.75in,footskip=0.75in]{geometry}

\usepackage{amsmath,amssymb}

\usepackage{changepage}

\usepackage[utf8x]{inputenc}

\usepackage{textcomp,marvosym}

\usepackage{cite}

\usepackage{nameref,hyperref}

\usepackage[right]{lineno}

\usepackage{microtype}
\DisableLigatures[f]{encoding = *, family = * }

\usepackage[table]{xcolor}

\usepackage{array}

\newcolumntype{+}{!{\vrule width 2pt}}

\newlength\savedwidth

\newcommand\thickhline{\noalign{\global\savedwidth\arrayrulewidth\global\arrayrulewidth 2pt}%
\hline
\noalign{\global\arrayrulewidth\savedwidth}}

\usepackage{ulem}

\raggedright
\usepackage[aboveskip=1pt,labelfont=bf,labelsep=period,justification=raggedright,singlelinecheck=off]{caption}

\makeatletter
\renewcommand{\@biblabel}[1]{\quad#1.}
\makeatother

\date{}

\usepackage{lastpage,fancyhdr,graphicx}
\usepackage{epstopdf}
\fancyheadoffset[L]{2.25in}
\fancyfootoffset[L]{2.25in}

\begin{document}
\vspace*{0.2in}

\begin{flushleft}
{\Large
\textbf\newline{Exploratory and Confirmatory Factor Analyses of Religiosity: A Four-Factor Conceptual Model} 
}
\newline
\\
Carlos Miguel Lemos\textsuperscript{1*},
Ross Joseph Gore\textsuperscript{2},
F. LeRon Shults\textsuperscript{3,4}
\\
\bigskip
\textbf{1} Institute for Religion, Philosophy and History, University of Agder, Kristiansand, Norway
\\
\textbf{2} Virginia Modeling, Analysis and Simulation Center, Old Dominion University, Norfolk, VA, United States of
America
\\
\textbf{3} Institute for Global Development and Planning, University of Agder, Kristiansand, Norway
\\
\textbf{4} Center for Modeling Social Systems at NORCE, Kristiansand, Norway
\\

\bigskip

%
%





* carlos.lemos@uia.no

\end{flushleft}
\section*{Abstract}
We describe an exploratory and confirmatory factor analysis of the International Social Survey Programme Religion Cumulation (1991-1998-2008) data set. The exploratory factor analysis was performed using data from the first two waves (1991 and 1998), and led to the identification of four factors which we labeled ``Religious formation,'' ``Supernatural beliefs,'' ``Belief in God,'' and ``Religious practice.'' The confirmatory factor analysis was run using data from 2008, and led to results consistent with the four-factor structure obtained in the previous step. We also run a set of structural equation models in an attempt to find whether this procedure could suggest causality links among the three factors related to the respondents' current religiosity, given the known causal precedence of ``Religious formation.'' The two SEMs with best fit suggest that religious practice influences beliefs more than the latter influence practice, but no causal precedences can be inferred from this result.   


\section{Introduction}
Religion has traditionally played a central role in human societies, shaping both personal development and cultural change. The complexity and diversity of religious phenomena have led to long-standing debates among scholars about how to conceptualize and measure ``religiosity.'' The scientific study of these highly contested and extremely complicated phenomena calls for a multidisciplinary approach encompassing fields as diverse as cognitive science, evolutionary biology, moral psychology, anthropology, sociology, and political science \cite{Shults2014}.

Several researchers have utilized statistical analyses of large data sets based on large scale surveys such as the World Values Survey (WVS), the European Values Survey (EVS) or the International Social Survey Programme (ISSP). For example, Gervais and Najle \cite{gervais_learned_2015} used three variables in the WVS related to belief in God, religious upbringing and attendance to services to show, via a signal detection approach, that religious upbringing has a strong influence on later belief in God. Norris and Inglehart \cite{norris_sacred_2011} presented a conceptual scheme connecting a number of hypotheses for explaining secularization (\cite{norris_sacred_2011}, Fig. 1.1, page 15), which they define as the erosion of religious participation, values and beliefs. They classify the latter as ``indicators of religiosity'' (\cite{norris_sacred_2011}, Table 2.1, page 41). Their factor analysis in Table 7.1 of \cite{norris_sacred_2011} describes three factors of ``Work Ethic,'' but no attempt was made to carry out a corresponding analysis on the religiosity ``indicators.'' Inglehart and Welzel  \cite{inglehart_modernization_2005,inglehart_development_2008} used sets of selected variables in the WVS, EVS and ISSP to provide supporting evidence for their theories on secularization and the human development path (both of which are related to religion). More recently, Doebler has identified three ``dimensions'' of religion (believing, belonging, and attendance) based on a multilevel analysis of the EVS \cite{Doebler2014}.  

However, none of the previous works was specifically directed towards the quantitative description of the core factors of individual religiosity, using a sufficiently representative set of variables related to attitudes and opinions about religion. Moreover, none of these studies applied the methods of factor analysis and structural equation modeling (SEM) in a systematic way. For example, the three indicators of religiosity considered by Norris and Inglehart \cite{norris_sacred_2011} were not proven to be factors of religiosity derived from statistical methods. Another problem is that even when factor analyses are used in the scientific study of religion, scholars rarely present detailed descriptions of the criteria used for acceptance or rejection of items and factors, the sequence of the elimination process, or estimates of reliability and goodness of fit for the solutions obtained.
 
In this article, we describe exploratory and confirmatory factor analyses of 34 selected variables in the ISSP Religion Cumulation data set \cite{hollinger_international_2011}, which combines the results from three waves of questionnaires on religion for years 1991, 1998 and 2008. The work was performed in three stages. First, we ran an exploratory factor analysis (EFA) using the first two waves (1991 and 1998). In this way we identified  four factor that we labeled ``Supernatural beliefs,'' ``Religious formation,'' ``Belief in God,'' ``Religious practice'' and ``Religious formation,'' the first three related to the respondent's current religiosity. Then, we used the last wave (2008) to run a confirmatory factor analysis (CFA), as EFA and CFA should be based on different data \cite{Hair2009}. This led to results consistent with those obtained in the first stage. Finally, we ran different structural equation models (SEMs) in an attempt to determine the causal relationships between the factors related to the current religiosity, taking into account the causal precedence of ``Religious formation.''

\section{Methods}

\subsection{Description of the ISSP Religion Cumulation Data Set. Selected Variables}
The ISSP Religion Cumulation data set\cite{hollinger_international_2011} contains the cumulated variables of the ISSP ``Religion'' surveys of 1991, 1998 and 2008 and comes in two separate files: a main file (ZA5070) with topic-related and background variables that appear in at least two surveys, and an add-on file (ZA5071) with variables that could not be cumulated for various reasons. The analysis in this article is based on the information in main file, which includes 122 variables (fields) for 102454 respondents from 28 countries. Only the countries that participated in at least two surveys (1998 and 2008) were included in the cumulation data file. For Germany and the United Kingdom two subsamples were included (East and West German, and Great Britain and Northern Ireland, respectively). Details on the contents, structure and coding of the ZA5070 cumulation file can be found in \cite{GESISDASS2011}.

Because of its size, the variables included and the proportion of valid respondents' answers, the ISSP Religion Cumulation data set is very valuable for extracting information about religious beliefs, practices and values. However, the ISSP ``Religion'' surveys have the important limitation that they mostly cover Christian Catholic, Protestant and Orthodox countries. Although Israel and Japan also contributed, there is no information from Muslim countries, India or China.

Table \ref{table1} shows the initial set of questions selected for the exploratory factor analysis as well as the variables' labels, number of valid levels and indication of score reversal. First, we identified the questions about religious beliefs, practices, and values that were most clearly related to the scientific study of religion, as well as others related to attitudes and values influenced by religiosity and confidence in churches and secular institutions. Finally, we also selected demographic variables that could help exploring other aspects often considered in the scientific study of religion such as age, gender and educational level. 

\begin{table}[!ht]
	\begin{adjustwidth}{-2.25in}{0in} 
		\centering
		\caption{
			{\bf Initial set of selected variables in the ISSP Religion Cumulation data set: variables' names, labels, number of valid levels and indication of polarity inversion.}}
		\begin{tabular}{|l+l|c|c|}
			\hline
			\multicolumn{1}{|l|}{\bf Variable} & 
			\multicolumn{1}{|l|}{\bf Question label} &
			\multicolumn{1}{|l|}{\bf \vbox{\hbox{Number of}\hbox{valid levels}}} &
			\multicolumn{1}{|l|}{\bf \vbox{\hbox{Score}\hbox{reversed}}} \\ \thickhline 
            \texttt{V11} & Sexual.relations.before.marriage & 4 & Yes\\ \hline
            \texttt{V12} & Sexual.relations.with.someone.other.than.spouse & 4 & Yes  \\ \hline
            \texttt{V13} & Sexual.relations.between.two.adults.of.the.same.sex & 4 & Yes \\ \hline
            \texttt{V14} & Opinion..Abortion.if.defect.in.the.baby & 4 & Yes \\ \hline
            \texttt{V15} & Opinion..Abortion.if.family.has.very.low.income & 4 & Yes \\ \hline
            \texttt{V16} & Husband.earn.money..wife.s.job.is.family & 5 & Yes \\ \hline
            \texttt{V20} & Confidence.in.parliament & 5 & Yes\\ \hline
            \texttt{V21} & Confidence.in.business.and.industry & 5 & Yes\\ \hline
            \texttt{V22} & Confidence.in.churches.and.religious.organizations & 5 & Yes\\ \hline
            \texttt{V23} & Confidence.in.courts.and.legal.system & 5 & Yes \\ \hline
            \texttt{V24} & Confidence.in.schools.and.educational.system & 5 & Yes \\ \hline
            \texttt{V25} & Religious.leaders.should.not.influence.vote & 5 & No \\ \hline
            \texttt{V26} & Religious.leaders.should.not.influence.government & 5 & No \\ \hline
            \texttt{V27} & Power.of.churches.and.religious.organizations & 5 & No \\ \hline
            \texttt{V28} & Closest.to.Rs.belief.about.God & 6 &  No\\ \hline
            \texttt{V29} & Best.describes.beliefs.about.God & 4 & No  \\ \hline
            \texttt{V30} & Belief.in.life.after.death & 4 & Yes \\ \hline
            \texttt{V31} & Belief.in.heaven & 4 & Yes \\ \hline
            \texttt{V32} & Belief.in.hell & 4 & Yes \\ \hline
            \texttt{V33} & Belief.in.religious.miracles & 4 & Yes \\ \hline
            \texttt{V35} & God.concerns.Himself.with.human.beings & 5 & Yes \\ \hline
            \texttt{V37} & Life.meaningful.because.God.exists & 5 & Yes \\ \hline
            \texttt{V46} & R.child..mother.attend.church & 9\textsuperscript{*} & No \\ \hline
            \texttt{V47} & R.child..father.attend.church & 9\textsuperscript{**} & No \\ \hline
            \texttt{V48} & R.age.11.12..R.attend.church & 9 & No \\ \hline
            \texttt{V49} & How.often.R.pray & 11 & No  \\ \hline
            \texttt{V50} & Take.part.in.church.activities & 11 & No \\ \hline
            \texttt{V51} & R.describes.self.as.religious & 7 & Yes \\ \hline
            \texttt{V64} & Modern.science.does.more.harm.than.good & 5 & Yes \\ \hline
            \texttt{V65} & Too.much.trust.in.science & 5 & Yes \\ \hline
            \texttt{ATTEND} & Attendance.of.religious.services & 6 & Yes \\ \hline
            \texttt{AGE} & Age.of.respondent & - & - \\ \hline
            \texttt{SEX} & Sex.of.respondent & 2 & - \\ \hline
            \texttt{DEGREE} & Highest.education.level.degree & 6 & No \\ \hline
		\end{tabular}
		\begin{flushleft} \textsuperscript{*} The uppermost level ``No mother/mother not present'' was merged with the lowest level ``Never.'' \\
			\textsuperscript{**} The uppermost level ``No father/father not present'' was merged with the lowest level ``Never.''
		\end{flushleft}
		\label{table1}
	\end{adjustwidth}
\end{table}

\subsection{Software Tools}
The data processing was done using \texttt{R} \cite{RCT2016}. Functions were written to import the \texttt{ZA5070\_v1-0-0.dta} Stata file and generate a R data frame, and to perform frequent queries and operations (retrieving a set of fields corresponding to selected questions, the records for a particular year and country or set of countries, etc.).

Exploratory factor analysis was done using the \texttt{psych} package \cite{Revelle2016}. Confirmatory factor analysis of the four reliable and strongly correlated factors of religiosity found in the exploratory factor analysis was done using \texttt{lavaan} \cite{Rosseel2012}. The structural equation modeling (SEM) analysis was done using \texttt{lavaan} as well.

\subsection{Data Preparation}
The R data frame generated from the \texttt{ZA5070\_v1-0-0.dta} Stata file was split in two data frames, one including the data from years 1991 and 1998 with 61928 records and another the data from year 2008 with 40526 records, so as to run EFA and CFA with distinct data.

The data preparation was done in the following steps (identical for EFA, CFA and SEMs).

First the selected variables were converted to ordered factors except \texttt{AGE} and \texttt{SEX} which were converted to numeric and unordered factor (categorical, nominal) respectively. Then, all ordinal variables were inspected to mark invalid/inconclusive answers as \texttt{NA} and remove spurious factor levels (e.g. levels for which no answer was recorded. After that, the ordinal variables were further inspected for score reversal. Table \ref{table1} shows the questions for which the scores were reversed.

In the EFA stage we computed the correlation matrices using three different methods via the \texttt{psych::mixed.cor} function: Pearson (simplest), Spearman, and mixed Spearman/polychoric with polychoric correlations being used for variables with up to six levels (inclusive). It was found that the differences between the correlation matrices computed using the three methods were negligible and Pearson correlation produced marginally better factor solutions.Therefore, all variables were converted to numeric in a final stage, which allowed faster and more efficient computation of EFA, CFA and SEMs.

\subsection{Exploratory Factor Analysis Method}
The EFA was done in two stages. In the first stage we computed factor solutions and eliminated variables (items) with low communality and low or cross-loadings \cite{Hair2009}, until all variables met the acceptance criteria described below, and a statistically significant and theoretically meaningful solution was found. This process was iterative, for removing variables also led to changes in the optimal number of factors to be extracted and consequently the variables' communality and loadings. In the second stage, the factors' reliability was assessed using the criteria described below, and a factor solution was computed using only the variables loading on the factors deemed reliable. This final factor solution was found to meet all acceptance criteria and to be theoretically meaningful, and the ensuing factor structure was then considered in the CFA and SEM analyses. 
 
The first stage consisted of the following steps:
\begin{enumerate}
	\item Inspection of the correlation structure and estimation of the optimal number of factors to be extracted;
	\item Computation of the factor analysis solution for the recommended number of factors as well as for one more and one less factor than recommended (i.e. ``bracketing'' on the number of factors) \cite{Hair2009};
	\item Inspection of the factor solution for variables with low communality and weak or cross loadings \cite{osborne_best_2009}, \cite{Hair2009};
	\item Elimination of variables that did not meet the acceptance criteria, if any, and re-doing the previous steps; otherwise, proceed to the second stage (inspection of factors' reliability).
\end{enumerate}

To determine the recommended number of factors, we used the scree test, the Very Simple Structure (VSS) and Velicer's Minimum Average Partial (MAP) criteria, as well as parallel analysis using \texttt{psych::fa.paralell} \cite{Revelle2016}. Parallel analysis consistently led to better estimates for the number of factors than the other methods, and was therefore the method of choice in the present work (bracketing on the number of factors as described in the second step above also confirmed this finding).

The acceptance criteria for the variables were: \textit{i}) communality $ > 0.40 $ \cite{osborne_best_2009} (reference \cite{Hair2009} recommends  $ > 0.50 $); \textit{ii}) loadings with absolute value $ > 0.32 $ (accounts for at least 10\% variance); and \textit{iii}) no cross loadings (variables with loadings with absolute value at 0.32 or higher on two or more factors \cite{osborne_best_2009}).

In the second stage, the factors obtained at the end of the first stage were inspected for reliability, using Cronbach's $ \alpha $ \cite{Cronbach1951} computed via the \texttt{psych::alpha} function. The acceptance criteria for factors were $ \alpha > 0.70 $ and at least three variables (items) loading on them \cite{Hair2009}. This last restriction is important for the subsequent CFA. Factors that failed to meet these criteria were removed from the analysis by eliminating the variables loading on them, and a final factor solution was computed and checked against all acceptance criteria (for variables and factors).

We now describe further details necessary for the replication of the EFA reported herein. The factor solutions were computed using \texttt{psych::fa} using data frames as inputs (instead of correlation matrices). Correlations were estimated using the Pearson coefficient (\texttt{cor = ``cor''}), as noted in the comment above on data preparation, with pairwise complete observations and no imputation of missing values. Oblique rotations (using the oblimin method, default in \texttt{psych::fa}) were used, since strong correlations between factors were to be expected. The factor extraction was done using the minimum residual method \cite{Comrey1962,Zegers1983}(default in \texttt{psych::fa}). We used one hundred bootstrap iterations and specified a maximum of 2000 iterations for convergence. The $ \alpha $ level for confidence intervals of loadings and factor correlations was set to 0.01. The remaining input parameters required by \texttt{psych::fa} were set to default values. 

\subsection{Confirmatory Factor Analysis Method}
We used CFA to test the measurement and model of the four factors suggested in EFA, i.e. to determine how well the measured variables represent the factors, and also to inspect the correlation structure among the factors. We used the 2008 data to run the CFA, since EFA and CFA should be done using independent data sets \cite{Hair2009}.

The CFA was done by first preparing the data as done for EFA. Then, the measurement and structural models of the final four-factor solution obtained in EFA were set up in and run in ``lavaan'' using \texttt{lavaan::cfa}. Following the standard procedures in CFA, no cross loadings were permitted (measured variables were forced to load on a single factor), and non null correlations between factors were allowed. The GOF of the confirmatory model was assessed using several different estimators (as described below) and the acceptance criteria were \cite{Hair2009}: \textit{i}) standardized loadings with (absolute) value $ > 0.50 $; \textit{ii}) average variance extracted (AVE) $ > 0.50 $; \textit{iii}) reliability of constructs (factors) $ > 0.70 $; and \textit{iv}) standardized residuals of measured variables $ < 2.50 $.

The details necessary for the replication of our CFA are as follows. The \texttt{lavaan::cfa} function was ran using a data frame of the variables loading on the four factors obtained in EFA with the records for ISSP  Religion 2008. The model was specified so that each variable (item) loaded on a single factor. The scale of the latent variables was set by fixing the factor loading of the first indicator to one (default in \texttt{lavaan::cfa}). The estimator used was diagonally weighted least squares (DWLS) \cite{Baghdarnia2014}. One hundred bootstrap iterations were used, as was done in EFA. The remaining parameters required to run \texttt{lavaan::cfa} were set to their default values.

\subsection{Structural Equation Modeling}
Next, we used structural equation modeling (SEM) to organize the relationships among the four factors obtained in EFA, taking into account that ``Religious formation'' has causal precedence over the other three factors. SEM enables us to specify path diagrams that hypothesize relationships among the four factors. Each diagram can be converted into a model with two components \cite{Rosseel2012}: 
\begin{enumerate}
  \item \textit{a measurement model} that describes the relationship of factors and their indicators (survey questions);
  \item \textit{structural equations} that depict regressions among factors as causal paths from one factor to another.
\end{enumerate}

Given a model, the structural equation modeling algorithm (\texttt{lavaan::sem}) computes the values for parameters within the measurement model and structural equation model that best match the observed data.  The best parameters were estimated using Full Maximum Likelihood Estimation (FML).

The ${\chi}^2$ is the statistical fit index used for assessing the GOF between the observed and model-implied covariance structures for the best solution for the model parameters. A ${\chi}^2$ that \textit{is not} significant implies a model with good fit. However, achieving a ${\chi}^2$  that is not significant for any model, regardless of fit quality, becomes increasingly difficult as the size of the sample increases. Most SEMs are based on samples of size $\tilde 200$. However, the sample size in our data set is measured in tens of thousands. Given the size of our data set, identifying a  ${\chi}^2$  that is not significant is exceptionally unlikely. This is not uncommon. In situations where the sample size is large and the ${\chi}^2$ tests lead to over-rejection, it is usual to evaluate the models' fit using alternative fit indices (AFIs), which belong to two categories: (1) absolute fit indices and (2) incremental fit indices \cite{hu1999cutoff, steiger2007understanding, bentler2007tests, iacobucci2010structural}.

Absolute AFIs are based on the discrepancy between the observed and model-implied covariance structure, and thus should be sufficiently small for a model to have good fit. We used two absolute  fit indices, namely the \textit{Root Mean Squared Error of Approximation} (RMSEA) and the \textit{Standardized Root Mean Square Residual} (SRMR). RMSEA is the standardized difference between the observed data and the SEM predicted data, while SRMR is defined as the standardized difference between the correlation of the observed data and the predictions of the SEM. Values of SRMR and RMSEA $<$ 0.08 reflect models with acceptable fit \cite{hu1999cutoff}. Incremental AFIs compare the fit of a given SEM with that of a null (or independence) model that only estimates the means and variances of the indicators (observed variables). The values of incremental AFIs of a SEM range from 0 (no better than the null model) to 1 (corresponding to a saturated model that predicts the observed variance-covariance matrix exactly). The two incremental fit measures employed in the evaluation of SEMs are the \textit{Comparative Fit Index} (CFI) and \textit{Tucker-Lewis Index} (TLI). Both apply penalties for including additional parameters in the model, but the TLI applies a more severe penalty than the CFI. Values of CFI and TLI $>$ 0.95 reflect models with acceptable fit \cite{hu1999cutoff}. The TLI is also referred to as \textit{Non-Normed Fit Index} or (NNFI).

The details necessary for the replication of our SEM are as follows. The SEM was done by first preparing the data as done for EFA and CFA. Then 588 different candidate models were generated. Each model was fit using a data frame of the variables, loading on the four factors obtained in EFA, with the records for ISSP  Religion 2008. The \texttt{lavaan::sem} function was run on the data with all function parameters set to their default values. The quality of fit was assessed using the four fit indices previously mentioned: CFI, TLI, RMSEA and SRMR. 

\section{Results}

\subsection{Exploratory Factor Analysis}
As outlined above, EFA was done in two stages. The first stage started by computing factor solutions using all variables in table \ref{table1} for a recommended (and best) twelve factors. Examination of this solution according to the criteria outlined above led to the exclusion of variables ``\texttt{V12},'' ``\texttt{V16},''  ``\texttt{V21},'' ``\texttt{V24},'' ``\texttt{AGE},'' ``\texttt{SEX}'' and ``\texttt{DEGREE}.'' In a second iteration, factor solutions were computed excluding these variables for a recommended (and best) ten factors. This led to a third iteration in which variables ``\texttt{V35},'' ``\texttt{V37},'' and ``\texttt{V64}'' were excluded and factor solutions were computed for a recommended (and best) eight factors. Examination of the best eight-factor solution led to further elimination of variables ``\texttt{V27}'' and ``\texttt{V65},'' the first due to low communality and the second because it did not load on any factor. The final iteration in stage one consisted of computing factor solutions for a recommended eight factors, which led to the best fit indexes.  
 
\begin{figure}[!h]
	\begin{adjustwidth}{-2.25in}{0in}
	\includegraphics[width = 1.0\linewidth]{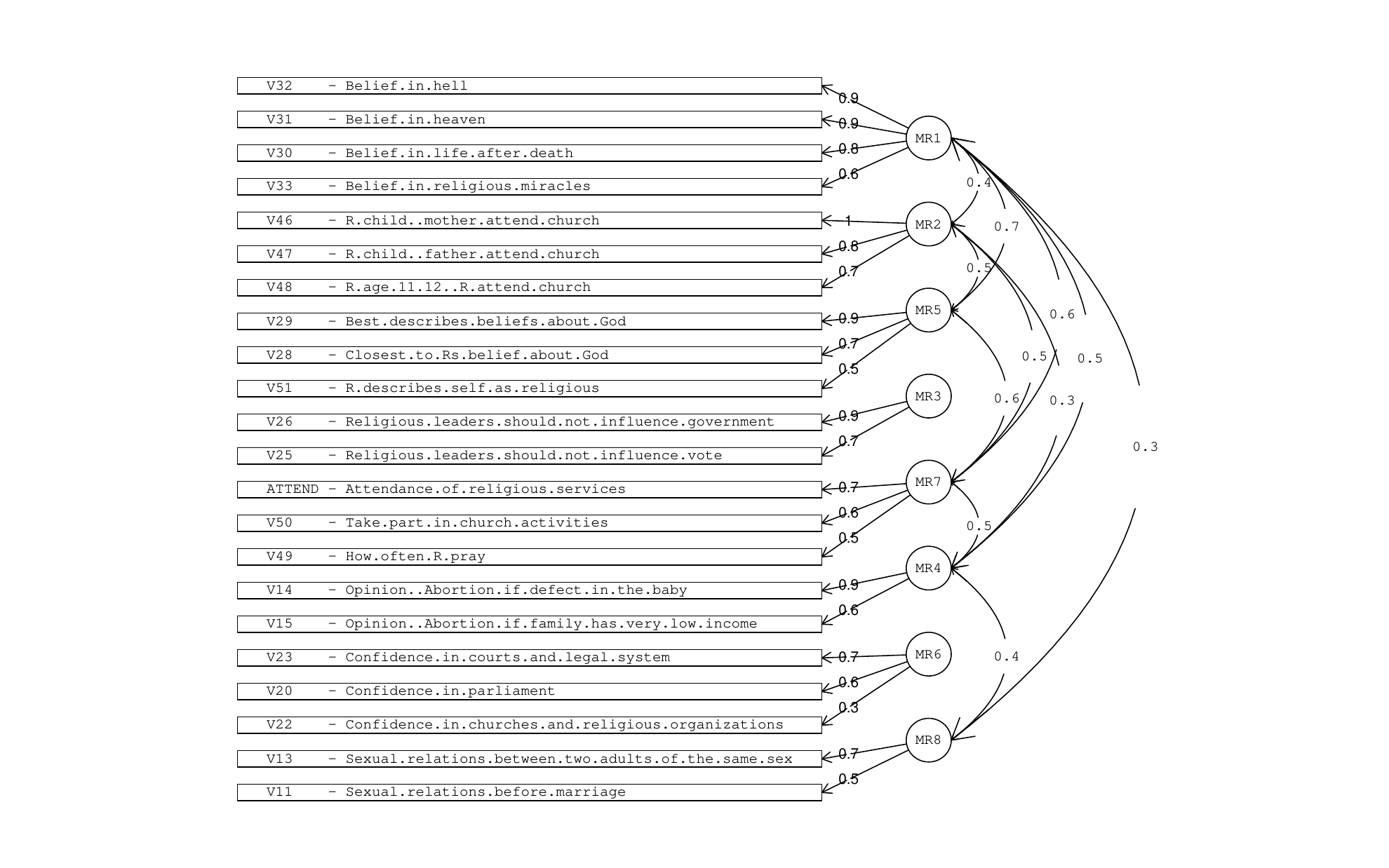}
		\caption{{\bf Factor solution diagram for the eight-factor solution.}
			Factor solution diagram for the eight-factor solution obtained in the EFA after the successive elimination of variables \texttt{V12}, \texttt{V16}, \texttt{V21}, \texttt{V24}, \texttt{V27}, \texttt{V35}, \texttt{V37}, \texttt{V64}, \texttt{V65}, \texttt{AGE}, \texttt{SEX} and \texttt{DEGREE}, drawn using \texttt{psych::fa.diagram}. Standardized item loadings and correlation coefficients between factors lower than 0.32 (10\% variance) are not represented.}
		\label{fig1}
	\end{adjustwidth}
\end{figure}

Fig \ref{fig1} shows the factor solution diagram for the eight-factor solution obtained at the final iteration of the first stage of the EFA. Table \ref{table2} shows the reliability estimates for the eight factors for several different estimators, computed using \texttt{psych::alpha}.
\begin{table}[!ht]
		\centering
		\caption{
			{\bf Reliability estimates for the eight-factor solution obtained in the first stage of EFA.}}
		\begin{tabular}{|l+l|l|l|l|l|l|l|l|}
			\hline
			\multicolumn{1}{|l|}{\bf } &
			\multicolumn{1}{|l|}{\bf raw $ \alpha $} & 
			\multicolumn{1}{|l|}{\bf std. $ \alpha $} &
			\multicolumn{1}{|l|}{\bf $ \lambda_{6} $} &
			\multicolumn{1}{|l|}{\bf average{\textunderscore}r} & 
			\multicolumn{1}{|l|}{\bf S/N} & 
			\multicolumn{1}{|l|}{\bf ase} & 
			\multicolumn{1}{|l|}{\bf mean} & 
			\multicolumn{1}{|l|}{\bf sd} \\ \thickhline 
			\texttt{MR1} &  0.90 &  0.90 &  0.88 & 0.70 & 9.1 & 0.00065 & 2.5 & 1.0 \\ \hline
			\texttt{MR2} &  0.87 &  0.87 &  0.82 & 0.69 & 6.5 & 0.00094 & 4.9 & 2.5 \\ \hline
			\texttt{MR3} &  0.78 &  0.78 &  0.64 & 0.64 & 3.6 & 0.0017  & 2.1 & 1.1 \\ \hline
			\texttt{MR4} &  0.73 &  0.74 &  0.58 & 0.58 & 2.8 & 0.0021  & 2.1 & 1.1 \\ \hline
			\texttt{MR5} &  0.89 &  0.90 &  0.87 & 0.75 & 9.2 & 0.00075 & 3.8 & 1.4 \\ \hline
			\texttt{MR6} &  0.56 &  0.57 &  0.48 & 0.31 & 1.3 & 0.0031  & 2.9 & 0.82 \\ \hline
			\texttt{MR7} &  0.75 &  0.81 &  0.75 & 0.58 & 4.2 & 0.0015  & 3.5 & 2.3 \\ \hline
			\texttt{MR8} &  0.62 &  0.63 &  0.46 & 0.46 & 1.7 & 0.003   & 2.3 & 1.1 \\ \hline
		\end{tabular}
		\begin{flushleft} \textbf{Description:} \textbf{raw $ \alpha $} is Cronbach's estimate based on covariances; \textbf{std. $ \alpha $} is Cronbach's estimate based on correlations; \textbf{$ \lambda_{6} $} is Guttman's $ \lambda_{6} $ reliability coefficient; \textbf{average{\textunderscore}r} is the average inter item correlation; \textbf{S/N} is the signal-to-noise ratio (or ratio of reliable variance to unreliable variance); \textbf{ase} is $ \alpha $'s standard error; \textbf{mean} is the mean of the summed scale constructed from the items; and \textbf{sd} is the standard deviation of the total factor scores \cite{Revelle2016}.
		\end{flushleft}
		\label{table2}
\end{table}

All factors in Fig. \ref{fig1} are seen to have a clear meaning. MR1 can be labeled ``Supernatural beliefs;'' MR2 ``Religious formation;'' MR3 ``Religion and politics'' (influence of religious leaders on vote and government); MR5 ``Belief in God;'' MR6 ``Confidence in institutions;'' MR7 ``Religious practice;'' and the two factors MR4 and MR8 are related to ``traditional values'' (or ``Conservatism'') with respect to opinions about abortion and sex, respectively. 

Inspection of table \ref{table2} shows that only factors MR1, MR2, MR5 and MR7 have $ \alpha > 0.7 $ and at least three variables loading on them, and therefore verify all acceptance criteria stated above. Also, they are also the ones with top ratio of reliable to unreliable variance (\textbf{S/N}). It should be noted that the factor MR2 (``Religious formation'') is associated to Christian religion, in which attendance at regular church services involves both men and women (contrary to the cas of Islam). also, the correlation between this factor and MR1, MR5 and MR7 is due to causality. MR1, MR5 and MR7 can be interpreted as ``core'' factors of the respondent's current religiosity. Although the diagram in Fig.1 is interesting for illustrating the correlation between religiosity and attitudes and opinions about traditional values, confidence in institutions and the relationship between religion and politics, we will not further address these aspects, and from now on concentrate on the factors associated with current religiosity and early religious socialization (MR", ``Religious formation'').

\begin{figure}[!h]
	\begin{adjustwidth}{-2.25in}{0in}
	\includegraphics[clip, trim=2cm 2cm 2.5cm 2cm,width = 1.0\linewidth]{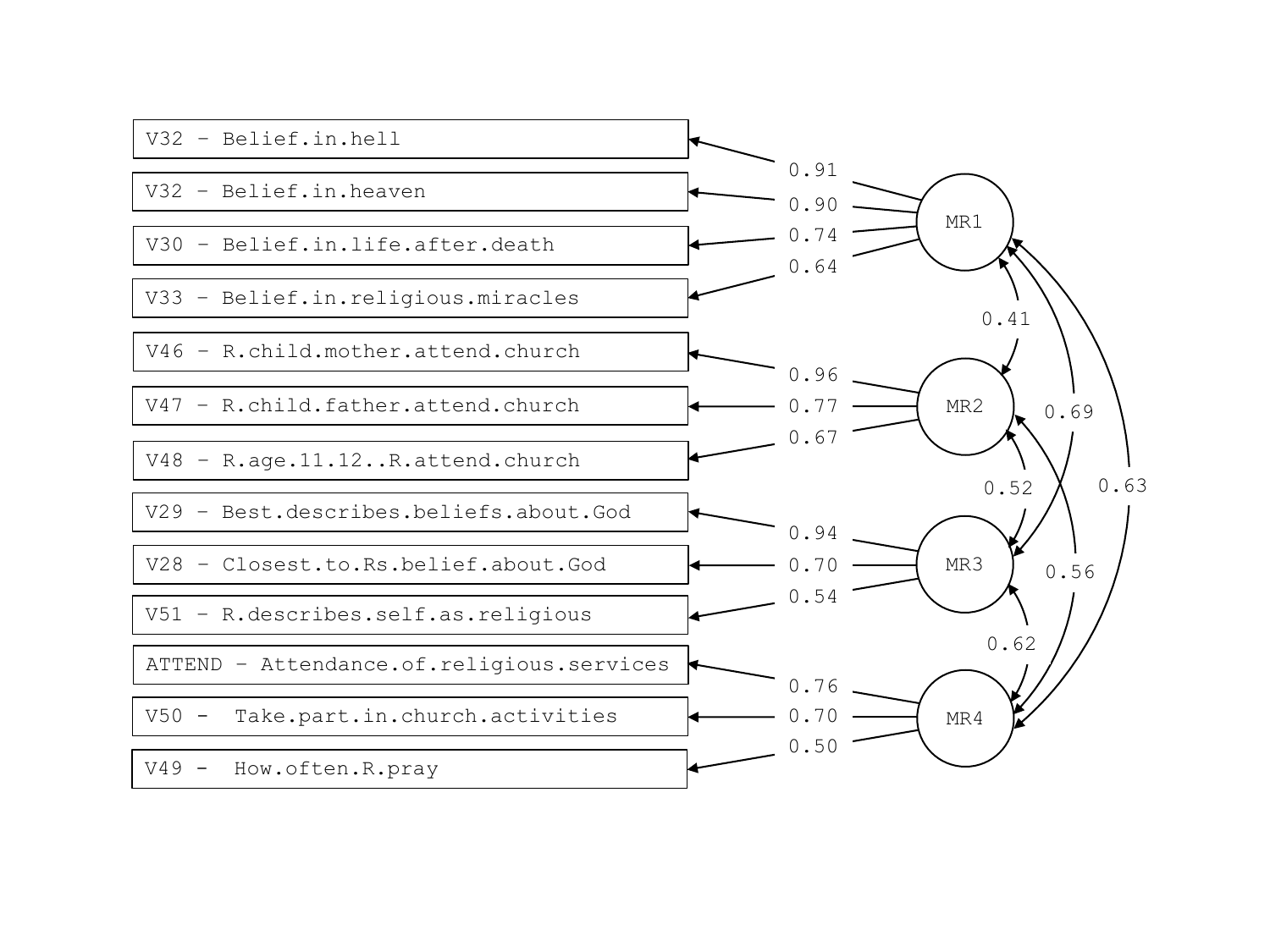}
		\caption{{\bf Factor Diagram for the four factors of religiosity.}
			Factor solution diagram for the four factors of religiosity obtained at the final stage of EFA, using \texttt{psych::fa.parallel}. Standardized item loadings and correlation coefficients between factors lower than 0.32 (10\% variance) are not represented. All loadings and correlations are positive due to the polarity inversions described in table \ref{table1}. Factor labels: MR1 $ \Longleftrightarrow $ ``Supernatural beliefs;'' MR2 $ \Longleftrightarrow $ ``Religious formation;'' MR3 $ \Longleftrightarrow $ ``Belief in God;'' MR4 $ \Longleftrightarrow $ ``Religious practice.''}
		\label{fig2}
		\begin{flushleft} 
			\textbf{Additional information:} Mean item complexity = 1.2; Tucker-Lewis Index (TLI) of factoring reliability = 0.99; Root means square error of approximation (RMSEA) index = 0.034 with 90\% confidence interval (0.033;0.035).
		\end{flushleft}
	\end{adjustwidth}
\end{figure}
\begin{table}[!ht]
	\centering
	\caption{
		{\bf Standardized factor loadings (pattern matrix), communality (h2), uniqueness (u2) and Hoffmann's item complexity (com) (see e.g. \cite{Hofmann1977,Pettersson2010}) for the four-factor model of individual religiosity.}}
	\begin{tabular}{|l+r|r|r|r|r|r|r|}
		\hline
		\multicolumn{1}{|l|}{\bf } &
		\multicolumn{1}{|r|}{\bf MR1} & 
		\multicolumn{1}{|r|}{\bf MR2} &
		\multicolumn{1}{|r|}{\bf MR3} &
		\multicolumn{1}{|r|}{\bf MR4} & 
		\multicolumn{1}{|r|}{\bf h2} & 
		\multicolumn{1}{|r|}{\bf u2} & 
		\multicolumn{1}{|r|}{\bf com}  \\ \thickhline 
        \texttt{V30}  &   0.74 & -0.03 & 0.07 & 0.01 & 0.61   & 0.39 & 1.0 \\ \hline
        \texttt{V31}  &   0.90 & 0.01  & 0.05 &-0.01 & 0.88   & 0.12 & 1.0 \\ \hline
        \texttt{V32}  &   0.91 & 0.03 & -0.09 & 0.00 & 0.74   & 0.26 & 1.0 \\ \hline
        \texttt{V33}  &   0.64 & 0.00 &  0.10 & 0.07 & 0.57   & 0.43 & 1.1 \\ \hline
        \texttt{V46}  &   0.02 & 0.96 & -0.02 & -0.05 & 0.87  & 0.13 & 1.0 \\ \hline
        \texttt{V47}  &   0.04 & 0.77 & -0.04 & 0.07 & 0.65   & 0.35 & 1.2 \\ \hline
        \texttt{V48}  &  -0.08 & 0.67 & 0.18  & 0.10 & 0.64 & 0.36 & 1.2 \\ \hline
        \texttt{V28}  &   0.16 & 0.00 &  0.70 & 0.09 & 0.77   & 0.23 & 1.1 \\ \hline
        \texttt{V29}  &   0.01 & 0.05 &  0.94 & -0.05 & 0.89  & 0.11 & 1.0 \\ \hline
        \texttt{V51}  &   0.05 & 0.01 &  0.54 &0.30  & 0.65   & 0.35 & 1.6 \\ \hline
        \texttt{V49}  &   0.11 & 0.05 &  0.31 & 0.50 & 0.70   & 0.30 & 1.8 \\ \hline
        \texttt{V50}  &   0.05 & -0.04 &  -0.09 & 0.70 & 0.44 & 0.56 & 1.0 \\ \hline
        \texttt{ATTEND} & 0.02 & 0.12 &  0.04 & 0.76 &  0.75  & 0.25 & 1.1 \\ \hline
	\end{tabular}
	\begin{flushleft} {\bf Factor labels: } MR1 $ \Longleftrightarrow $ ``Supernatural beliefs;'' MR2 $ \Longleftrightarrow $ ``Religious formation;'' MR3 $ \Longleftrightarrow $ ``Belief in God;'' MR4 $ \Longleftrightarrow $ ``Religious practice.'' 
	\end{flushleft}
	\label{table3}
\end{table}
\begin{table}[!ht]
	\centering
	\caption{
		{\bf Factor correlations and 99\% confidence intervals estimated via bootstrap iterations.}}
	\begin{tabular}{|l+r|r|r|}
		\hline
		\multicolumn{1}{|l|}{\bf } &
		\multicolumn{1}{|r|}{\bf lower} & 
		\multicolumn{1}{|r|}{\bf estimate} &
		\multicolumn{1}{|r|}{\bf upper}  \\ \thickhline 
	    MR1-MR2 & 0.40 & 0.41 & 0.42 \\ \hline 
	    MR1-MR3 & 0.68 & 0.69 & 0.69 \\ \hline 
	    MR1-MR4 & 0.62 & 0.63 & 0.64 \\ \hline 
	    MR2-MR3 & 0.51 & 0.52 & 0.53 \\ \hline 
  	    MR2-MR4 & 0.55 & 0.56 & 0.57 \\ \hline 
	    MR3-MR4 & 0.61 & 0.62 & 0.63 \\ \hline 	
	\end{tabular}
	\begin{flushleft} {\bf Factor labels: } MR1 $ \Longleftrightarrow $ ``Supernatural beliefs;'' MR2 $ \Longleftrightarrow $ ``Religious formation;'' MR3 $ \Longleftrightarrow $ ``Belief in God;'' MR4 $ \Longleftrightarrow $ ``Religious practice.'' 
	\end{flushleft}
	\label{table4}
\end{table}

Fig. \ref{fig2} shows the diagram for the four-factor model of religiosity. Table \ref{table3} shows the standardized factor loadings, communality, uniqueness and Hoffmann complexity index \cite{Hofmann1977,Pettersson2010} for all variables, and table \ref{table4} shows the corresponding estimates and 99\% confidence intervals for the factor correlations. 

\subsection{Confirmatory Factor Analysis}
\begin{figure}[!h]
	\begin{adjustwidth}{-2.25in}{0in}
	\includegraphics[width = 1.0\linewidth]{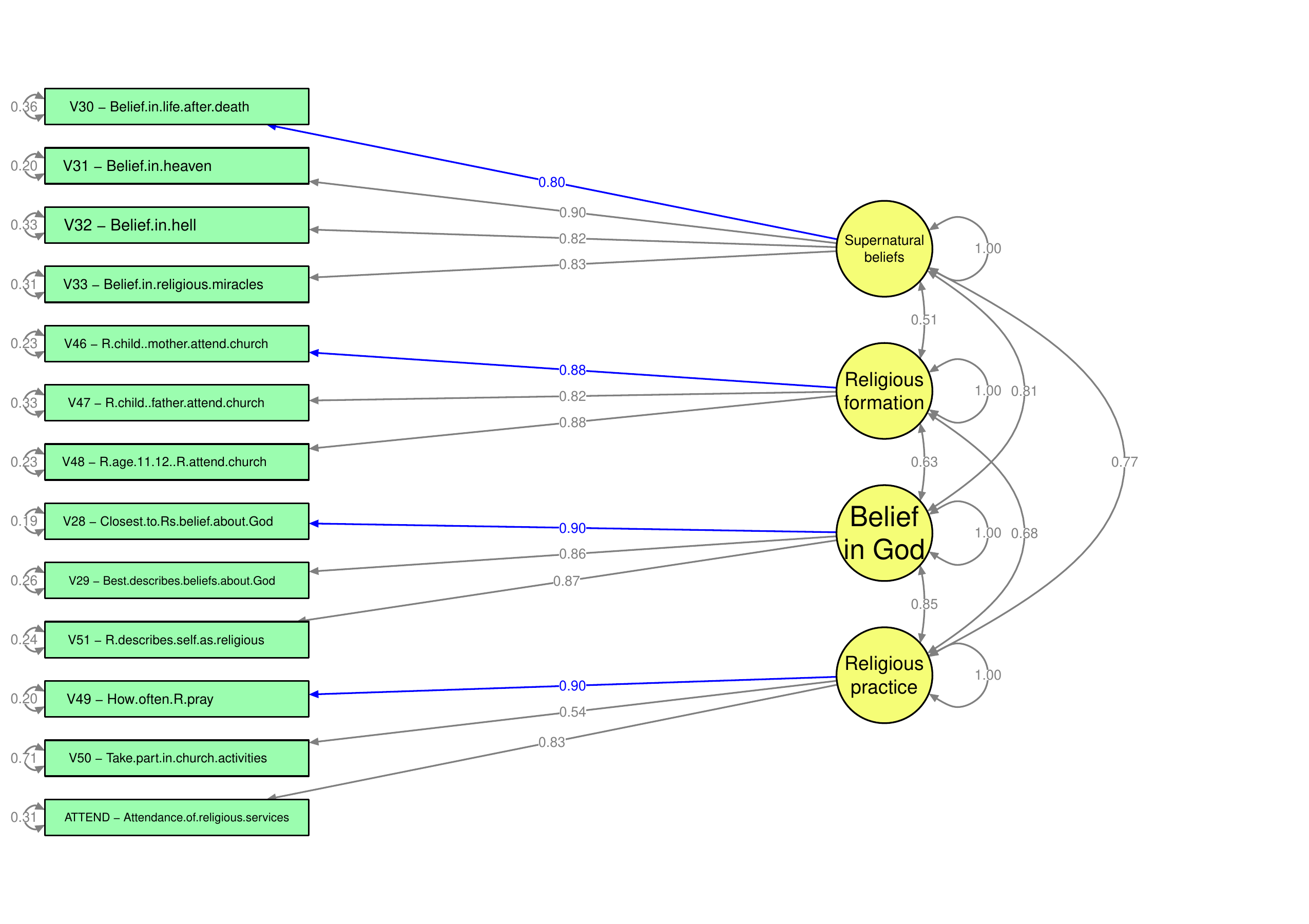}
		\caption{{\bf Path Diagram for CFA of the four factor model of religiosity.}
			Path diagram standardized parameter estimates for CFA of the four factor model of religiosity computed with \texttt{lavaan::cfa}, drawn using \texttt{semPlot::semPaths}. }
		\label{fig3}
		\begin{flushleft} 
			\textbf{Additional information:}   Test statistic $ chi_{2} = 1773.56 $ with $ df = 59 $, p-value 0; Comparative fit index (CFI) = 0.997; TLI = 0.997; RMSEA =  0.035; Normed Fit Index (NFI) = 0.997; Bentler-Bonett Nonnormed Fit Index (NNFI) = 0.997.
		\end{flushleft}
	\end{adjustwidth}
\end{figure}

Fig. \ref{fig3} shows the path diagram of the standardized parameter estimates obtained in CFA of a model built upon the four factors of religiosity identified in EFA and the variables (items) loading on them, using the procedure and settings described in the ``Methods'' section above. The CFA was computed using \texttt{lavaan:cfa}, which converged after 76 iterations, using 23636 out of 40526 observations.

\begin{table}[!ht]
	\centering
	\caption{
		{\bf Reliability estimates for the CFA model computed using the \texttt{semTools::reliability} function.}}
	\begin{tabular}{|l+r|r|r|r|}
		\hline
		\multicolumn{1}{|l|}{\bf } &
		\multicolumn{1}{|r|}{\bf \vbox{\hbox{Supernatural}\hbox{beliefs}} } & 
		\multicolumn{1}{|r|}{\bf \vbox{\hbox{Religious}\hbox{formation}} } &
		\multicolumn{1}{|r|}{\bf \vbox{\hbox{Belief}\hbox{in God}} } &
		\multicolumn{1}{|r|}{\bf \vbox{\hbox{Religious}\hbox{practice}} }   \\ \thickhline 
		$ \alpha $  &  0.903  &   0.891  &   0.892   &  0.774  \\ \hline
		$ \omega_{1} $  &  0.904  &   0.892  &   0.912   &  0.845  \\ \hline
		$ \omega_{2} $ & 0.904  &   0.892  &   0.912   &  0.845  \\ \hline
		$ \omega_{3} $ & 0.905  &   0.893  &   0.916   &  0.838  \\ \hline
		AVE    &   0.703     &      0.734  &   0.781   &  0.682   \\ \hline
	\end{tabular}
	\begin{flushleft} {\bf Additional information: } $ \alpha $ is Cronbach's reliability coefficient. The theoretical description of the $ \omega_{1} $, $ \omega_{2} $ and $ \omega_{3} $ coefficients can be found in \cite{Raykov2001}, \cite{Bentler2009} and \cite{McDonald1999} respectively.
	\end{flushleft}
	\label{table5}
\end{table}

Table \ref{table5} shows some reliability estimates for the CFA model. Combining the information presented in Fig. \ref{fig3} and table \ref{table5}, it can be observed that the factors' reliability and AVE meet the criteria stated the the ``Methods'' section. Therefore, the four-factor model was further studied using SEMs to investigate how the fit measures depend on other types of links (e.g. regression) between factors, in an attempt to determine whether the SEMs suggested any clue about eventual causal precedences between the factors related to the current religiosity, or to differences between the effects of practice on each of the beliefs factors.
 
\subsection{Structural Equation Modeling}
The four-factor model analysed in the previous sections shows that the four factors are correlated, and which items measure each of the factors. However, at the structural level, ``Religious formation'' is not related to the current religiosity and is a causal antecedent of the other three factors. Thus, we wanted to study whether the re-specification of the structural model, keeping the same measurement model, would change the fit measures and possibly suggest causal relations among the three factors related to the current religiosity.
  
Next, we applied structural equation modeling (SEM) to organize the relationships among the four clear and strongly correlated latent variables (MR1 - ``Supernatural beliefs;'' MR2 - ``Religious formation;'' MR3 - ``Belief in God;'' MR4 - ``Religious practice''). Given our four factors it is possible to construct 588 SEMs. As explained in the Methods section, we used four AFIs to assess the models' fit: CFI, TLI, SRMR and RMSEA.

\begin{figure}[!ht]
	\begin{adjustwidth}{-2.25in}{0in}
		\begin{center}
			\includegraphics[width=6in]{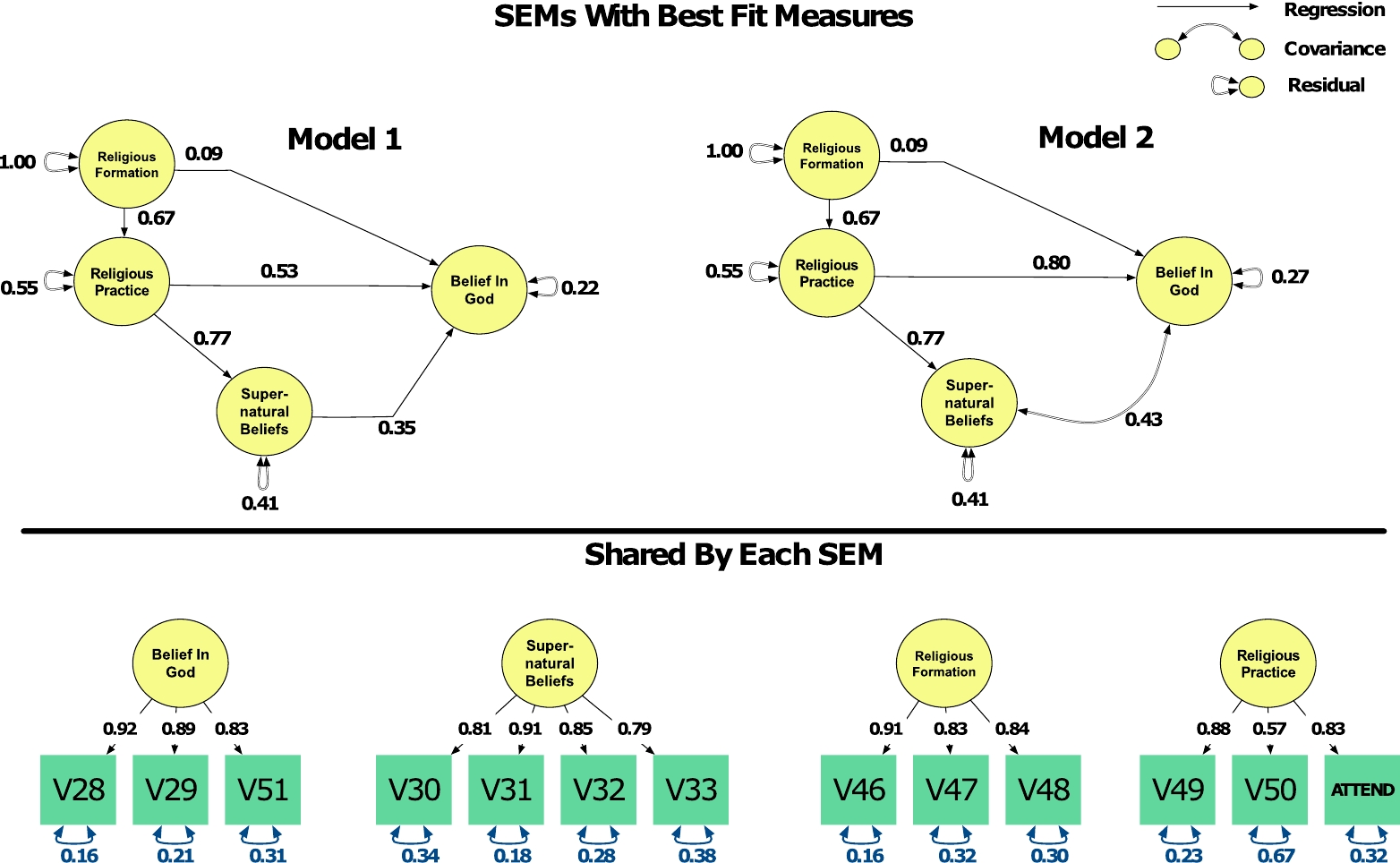}
		\end{center}
		\caption{\bf The two SEMs with the best measures for the four fit indices that are consistent with the causal precedence of the ``Religious formation'' relative to the other three factors (related to beliefs and practice).}
		\label{lab:DifferentSEMs}
	\end{adjustwidth}
\end{figure}

Of these 588 SEMs, the two models that are consistent with the causal precedence of the ``Religious formation'' factor and have the best fit measures are shown in Fig. \ref{lab:DifferentSEMs}. In addition, the models do not have invalid parameter values (negative residuals or variances or correlations larger than 1.0). Furthermore, the regression coefficients, factor loadings and covariances all have the expected sign (positive). 

\begin{table}[!ht]
	\centering
	\caption{
		{\bf Number of parameters, degrees of freedom, ${\chi}^2$ and fit indices for the two best fitting SEMs that are consistent with the causal precedence of the ``Religious formation'' relative to the other three factors (related to beliefs and practice).}}
	\begin{tabular}{|l+r|r|r|r|r|r|r|}
		\hline
		\multicolumn{1}{|l|}{\bf } &
		\multicolumn{1}{|r|}{\bf \# Pars} & 
		\multicolumn{1}{|r|}{\bf df} &
		\multicolumn{1}{|r|}{\bf ${\chi}^2$} &
		\multicolumn{1}{|r|}{\bf CFI} &
		\multicolumn{1}{|r|}{\bf TLI/NNFI}  & 
		\multicolumn{1}{|r|}{\bf RMSEA}  & 
		\multicolumn{1}{|r|}{\bf SRMR} \\ \thickhline 
	    Model 1 & 31 & 60 & 7729.068 & 0.968 & 0.958 & 0.074 & 0.034 \\ \hline 
	    Model 2 & 31 & 60 & 7729.068 & 0.968 & 0.958 & 0.074 & 0.034 \\ \hline  
	\end{tabular}
	\label{table:BestSemFit-Reg}
\end{table}

\begin{table}[!ht]
	\centering
	\caption{
		{\bf Additional fit indices for the two best fitting SEMs that are consistent with the causal precedence of the ``Religious formation'' relative to the other three factors (related to beliefs and practice). Upper and Lower RMSEA are based on 90\% confidence interval.}}
	\begin{tabular}{|l+r|c|c|}
		\hline
		\multicolumn{1}{|l|}{\bf } &
		\multicolumn{1}{|r|}{\bf NFI} &
		\multicolumn{1}{|r|}{\bf Lower RMSEA CI}  & 
		\multicolumn{1}{|r|}{\bf Upper RMSEA CI}  \\ \thickhline 
	    Model 1 & 0.967 & 0.072 & 0.075 \\ \hline 
	    Model 2 & 0.967 & 0.072 & 0.075  \\ \hline 
	\end{tabular}
	\label{table:BestSemFit-Extra}
\end{table}

The number of parameters (\# Pars), degrees of freedom (DF) and four fit indices for the two models are provided in Table \ref{table:BestSemFit-Reg}. Table \ref{table:BestSemFit-Extra} provides several additional fit indices for the models including the Bentler-Bonett Index or Normed Fit Index (NFI) and upper and lower confidence intervals for the RMSEA. Based on the evaluation measures shown in Table \ref{table:BestSemFit-Reg} and \ref{table:BestSemFit-Extra}, none of the two models is statistically superior to the other. However, both models suggest that religious practice influences beliefs more than the latter influence practice, but no causal precedences can be inferred from this result. 

\section{Discussion}
The EFA and CFA analyses of the ISSP Religion Cumulation data described herein led to the identification of four factors, one related to early religious socialization (``Religious formation'') and three to the respondent's current religiosity (``Supernatural beliefs,'' ``Belief in God'' and ``Religious practice'').

The existence of two distinct factors of religious beliefs can be explained by the fact that afterlife beliefs is related to death anxiety as postulated by thanatocentric theories \cite{Jong_2016}, whereas belief in God is culturally acquired and has different systemic and psychological roots \cite{Norenzayan2015}. The strong correlation was found between factors such as belief in the supernatural (and God) and attendance at religious services is not surprising. The relationship between the tendency to believe in disembodied intentional forces and the tendency to signal commitment to an in-group through ritual participation is extremely well documented in fields such as the cognitive science of religion and cultural anthropology  \cite{Atran2002,McCauley2002,Whitehouse2004,Martinez2013}.

The fit measures of the models shown in Fig. 4 are substantially worse that those of the model shown in Fig. 3. Thus, while the four factors seem to be reliably measured by the items suggested by the EFA, the levels of the three factors related to the current religiosity are significantly affected by factors other than family attendance at regular church services during the respondents' formative years (``Religious formation''), because the fit measures degraded when correlation paths were replaced by regression paths from the latter factor to the other three.

\subsection{Limitations}
Factor analysis is very sensitive to a large number of aspects, such as the existence of heterogeneous group structure(s) in the data (countries, religious groups, etc.), the way the covariance structure is computed (Pearson, Spearman, Kendall-$ \tau $, polychoric/tetrachoric/polyserial), the factor extraction methods in EFA, and the estimation method used in CFA.

While methods similar to the ones described herein have been used in several works that relate religiosity to indices that measure constructs such as existential security and human development \cite{inglehart_modernization_2005,inglehart_development_2008,norris_sacred_2011}, researchers must be aware that these methods have important limitations for cross-cultural analyses. In particular, they are not suitable for performing more rigorous statistical comparisons between different groups (e.g. nonreligious people and different Christian groups). This requires multi-group analysis techniques that are substantially more complicated than the ones used in this work \cite{Lemos2019}.

\section{Conclusion}
We performed exploratory and confirmatory factor analyses of religiosity based on ISSP Religion Cumulation data set (1991-1998-2008). The EFA began with 34 selected variables using the data for the 1991 and 1998 waves and led to the identification of four factors, ``Supernatural beliefs,'' ``Belief in God,'' ``Religious practice'' and ``Religious formation,'' the last of which is related to the frequency of family attendance at church services during the respondents' formative years and is causally precedent relative to the other three factors. The latter represent different aspects of the respondents' current religiosity, whose meaning and inter-relation had already been documented in the literature \cite{Atran2002,McCauley2002,Whitehouse2004,Martinez2013,Norenzayan2015,Jong_2016}.

The measurement model of the four (or ``$ 1 + 3 $'') factors suggested was tested using CFA based on the data for the 2008 wave. The resulting CFA model had good fit measures, suggesting that the items loading on each factor provide reliable measurements.

After the identification of the four factors using EFA and the confirmation of their measurement indicators using CFA, we performed an exploration of the four-factor model using SEMs, taking into account the causal precedence of the frequency of past family attendance (``Religious formation'') relative to the other three factors related to current religiosity, to investigate whether or not these would suggest any precedence among the latter (although SEM do not prove causation). We found that when correlation paths were replaced from regression paths from ``Religious formation'' the fit measures degraded substantially. Thus, the levels of current religiosity seem to be significantly affected by factors not considered in the SEMs. The two SEMs with best fit suggest that religious practice influences beliefs more than the latter influence practice, but no causal precedences can be inferred from this result. 

\section*{Supporting Information}

\paragraph*{ISSP ``Religion'' cumulation of the years 1991, 1998 and 2008.}
\label{S1_ISSP}
Available in STATA (\texttt{ZA5070\_v1-0.0.dta}) and SPSS (\texttt{ZA5070\_v1-0.0.sav}) formats from: \small{\url{https://dbk.gesis.org/dbksearch/sdesc2.asp?no=5070\&db=e\&doi=10.4232/1.10860}}.

\paragraph*{Guide for the ISSP ``Religion'' Cumulation file (\texttt{ZA5070\_v1-0.0)}.}
\label{S2_ISSP}
\normalsize{Guide for the ISSP ``Religion'' cumulation of the years 1991, 1998 and 2008 with the variable names and labels shown in table \ref{table1} is available from: \url{http://zacat.gesis.org/webview/index/en/ZACAT/ZACAT.c.ZACAT/ISSP.d.58/Cumulations.d.72/International-Social-Survey-Programme-Religion-I-III-ISSP-1991-1998-2008-/fStudy/ZA5070}.}

\section*{Acknowledgments}
Funding for this research was provided by the Research Council of Norway (grant \#250449).

We would like to thank Saikou Diallo, Associate Research Professor at Old Dominion University, for graciously offering us advice that helped hone our methodology and analysis in this work.

\nolinenumbers

\end{document}